\documentclass[12pt,a4paper,final]{iopart}
\expandafter\let\csname equation*\endcsname\relax
\expandafter\let\csname endequation*\endcsname\relax
\usepackage{iopams}
\usepackage{graphicx}
\usepackage{subfigure}
\usepackage[breaklinks=true,colorlinks=true,linkcolor=blue,urlcolor=blue,citecolor=blue]{hyperref}
\usepackage{cite}
\usepackage{braket}
\usepackage{mathtools}

\usepackage[a4paper]{geometry}
\usepackage{bpchem,upgreek}
	\usepackage{amsmath}
	\usepackage{makeidx}
	\usepackage{amsfonts}
	\usepackage[ansinew]{inputenc}
	\usepackage[usenames,dvipsnames]{pstricks}
    \usepackage{graphicx}
    \usepackage{float}
    \usepackage{subfigure}
	\usepackage{pst-grad} 
	\usepackage{pst-plot} 
	\usepackage{sidecap}
	\usepackage[makeroom]{cancel}

\begin{document}

\title[Teleportation of squeezed states...]{Teleportation of squeezed states in the absence and presence of dissipation}

\author{N Sehati$^{1}$}
\author{M K Tavassoly$^{1}$}
\address{$^1$Atomic and Molecular Group, Faculty of Physics, Yazd University, Yazd  89195-741, Iran}
\ead{mktavassoly@yazd.ac.ir}

\author{M Ghasemi$^{1}$}

\vspace{10pt}
\date{today}

\begin{abstract}
 In this paper at first we successfully teleport the unknown quantum state which is a superposition of squeezed vacuum state and squeezed one-photon state using the beam splitter in the absence of dissipation. In the continuation, we try to implement the same teleportation protocol, however, in the presence of dissipation effects. To do this task, we use proper entangled channel to reach to perfect teleportation under the influence of decoherence. Finally, we consider another superposition of two squeezed vacuum states with separation in phase by $\pi$ and teleport it with a different appropriate entangled channel. In fact, we will observe that, one can successfully teleport the considered superposition of squeezed states by choosing proper entangled channels in the presence and absence of dissipation in appropriate chosen conditions.
\end{abstract}

\pacs{03.65.Yz; 03.67.Bg; 42.50.-p; 42.79.Fm}

\vspace{2pc}
%
%
%
%

\section{Introduction}
 Teleportation of quantum states first introduced by Bennett et al. \cite{Bennett1993} is a very interesting and useful phenomenon in quantum information processing by which an unknown quantum state can be transmitted from a location to another \cite{Bouwmeester1997}. Quantum entanglement has an  important role in this phenomenon. In this respect, the quantum channel which is used for teleportation purpose should be entangled. These channels can be produced by using the beam splitter \cite{Agarwal2012}, entanglement swapping process \cite{Ghasemi2016,Ghasemi2017,Pakniat20172} and performing appropriate interaction performance \cite{Zou2003,Ghasemi2019,Ghasemi2018,Ghasemi20192,Baghshahi2015}.
  After preparation of proper quantum channel, quantum teleportation can be implemented.\\
 In this regard, generation of various classes of entangled states is of enough interest,  moreover, the entangled coherent states (quasi-Bell state) \cite{Paternostro2003,Gerry1997,Jeong2006} and special classes of entangled squeezed states \cite{Dibakar2015,Karimi2015} can be generated. The entangled squeezed states, will be used in this paper, have been generated via different schemes, \textit{i.e.}, in atomic Bose-Einstein condensates \cite{Kuang2003}, with the help of QED method \cite{Zheng2010,Karimi2015} and by utilizing the beam splitter \cite{Zhou2002}. Also, these states can be produced by, for instance, degenerate parametric down conversion in an optical cavity \cite{Wu1986}, in a resonator with a moving wall \cite{Dodonov1990} and via non-degenerate four-wave mixing in an optical cavity \cite{Slusher1985,Yurke1984}.  \\
   Quantum teleportation has been studied by considering different methods such as QED method \cite{Pakniat2017,Pakniat2016,Sehati2017} and by using beam splitter \cite{Van2001,Ralph1998,Takei2005,Xin2003}. Also, in \cite{Milburn1999} quantum teleportation using two-mode squeezed vacuum state has been investigated. Teleportation of quantum continuous variables \cite{Braunstein1998}, photon states \cite{Kim2001} and atoms \cite{Riebe2004,Barrett2004} have already been investigated. The authors in \cite{Zhang2000} have proposed a scheme to produce an EPR pair, by combining two bright amplitude squeezed lights and they have demonstrated that this protocol can be used to perform quantum teleportation of continuous variables. Teleportation of a quantum state using three-particle entanglement \cite{Karlsson1998} and nuclear magnetic resonance  have also been previously considered \cite{Nielsen1998}. The authors in \cite{Olmschenk2009} have demonstrated the quantum teleportation between distant matter qubits. Teleportation of coherent state using the beam splitter has been studied in \cite{Van2001}. To access the reliable and realizable teleportation scheme, the effects of decoherence on this scheme have also been considered \cite{Van2001,El2019,Man2012}.\\
    In present work, our protocol for teleportation, which is based on utilizing the beam splitter is introduced, by which we successfully teleport the unknown quantum state (which is a superposition of squeezed vacuum and one-photon squeezed state). Then, we consider the effects of decoherence in the teleportation scheme. This decoherence is modeled via mixing the light mode with the vacuum state at the beam splitter (see Refs. \cite{Van2001,Bergmann2017,Wickert2010}). In addition, teleportation of a superposition of two squeezed vacuum states with separation in phase by $180^\circ$ is considered.\\  
   The paper organizes as follows: We teleport successfully the unknown quantum state which is a superposition of a squeezed vacuum state and a squeezed one-photon state by using the beam splitter in the absence of dissipation in Sec. $2$. Then, in Sec. $3$, we consider the effect of decoherence and teleport prior unknown state. In the continuation of Sec. $3$ we pay our attention to the teleportation of another unknown quantum state which is a superposition of two squeezed vacuum states with separation in phase by $180^\circ$. Finally, the paper ends with a summary and conclusions in Sec. $4$.

\section{Teleportation of a superposition of squeezed states}\label{sec.model}

 \begin{figure}[H]
   \centering
 \includegraphics[width=0.75\textwidth] {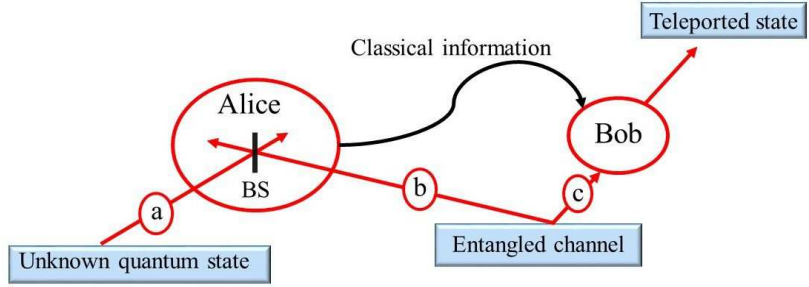}
   \caption{\label{fig:Fig1} {The scheme of quantum teleportation from Alice to Bob by using entangled channel via a beam splitter.}}
  \end{figure}
  In this section, we want to teleport the unknown quantum state
  \begingroup\makeatletter\def\f@size{9}\check@mathfonts
  \begin{eqnarray}\label{ateleportedunknownstate}
  \ket{\psi}_a=\epsilon_{+}\ket{\xi}_a+\epsilon_{-}\ket{\xi^{'}}_a,
   \end{eqnarray}
   where $ \epsilon_{\pm} $ are the undetermined coefficients that satisfy $\vert{\epsilon_{+}}\vert^{2}+\vert{\epsilon_{-}}\vert^{2}=1$. The state (\ref{ateleportedunknownstate}) includes all Fock states (odd and even basis), since $\ket{\xi}$ and $\ket{\xi^{'}}$ are the squeezed vacuum and one-photon states, respectively \cite{Quesne2001}, \textit{i.e.},
   \begin{eqnarray}\label{ unknown state}
   \ket{\xi}&=& S(\xi)\ket{0}=\Sigma^\infty_{n=0}{C_{2n}\ket{2n}},\\  \nonumber
   \ket{\xi^{'}}&=& S(\xi)\ket{1}=\Sigma^\infty_{n=0}{D_{2n+1}\ket{2n+1}},
   \end{eqnarray}
   where $\xi=r e^{i\varphi}$ and $S(\xi)=\exp{[\frac{1}{2}(\xi^{\ast}\hat{a}^{2}-\xi\hat{a}^{\dagger2})]}$ is the squeezing operator by which the expansion coefficients in (\ref{ unknown state}) result respectively in \cite{Nieto1997}
    \begin{eqnarray}\label{The coefficient of squeezed state}
    C_{2n}=\dfrac{1}{\sqrt{\cosh{r}}} \dfrac{\sqrt{(2n)!}}{2^{n}{n}!}e^{in\varphi}\tanh^{n}{r}, \\ \nonumber
    D_{2n+1}=\dfrac{1}{\cosh^{\frac{3}{2}}{r}} \dfrac{\sqrt{(2n+1)!}}{2^{n}{n}!}e^{in\varphi}\tanh^{n}{r}.
    \end{eqnarray}
   The teleportation protocol is as follows: consider a situation where a field mode b is stored in Alice's lab, while mode c describes a light beam propagating to Bob (see figure \ref{fig:Fig1}). Initially, Alice and Bob  share the maximally entangled channel,
   \begin{eqnarray}\label{channel}
   \ket{H_{\xi}}_{b,c}=\dfrac{1}{\sqrt{2}}\left( {\ket{-\xi}_{b}\ket{\xi^{'}}_{c}+\ket{-\xi^{'}}_{b}\ket{\xi}_{c}}\right),
   \end{eqnarray}
    which is made up of squeezed one-photon and vacuum states. So, the whole state of the system reads as: 
   \begin{eqnarray}\label{total state}
   \ket{\psi}_{a}\otimes\ket{H_{\xi}}_{b,c}&=&\dfrac{1}{\sqrt{2}}\left[ \epsilon_{+}\ket{\xi}_{a}\ket{-\xi}_{b}\ket{\xi^{'}}_{c}+\epsilon_{+}\ket{\xi}_{a}\ket{-\xi^{'}}_{b}\ket{\xi}_{c}\right. \\ \nonumber
   &+&\left. \epsilon_{-}\ket{\xi^{'}}_a\ket{-\xi}_{b}\ket{\xi^{'}}_{c}+\epsilon_{-}\ket{\xi^{'}}_a\ket{-\xi^{'}}_{b}\ket{\xi}_{c}\right] .
   \end{eqnarray}
   Now, to achieve the teleportation purpose, Alice mixes her part of entangled state (field mode b) with the state to be teleported (field mode a) using a beam splitter. Accordingly, the total state after the beam splitter can, in principle, be obtained via the relation,
   \begin{equation}
   \ket{\psi}_{\mathrm{out}} =u_{\mathrm{BS}}(\theta)\ket{\psi}_{a}\otimes\ket{H_{\xi}}_{b,c},
   \end{equation}
   where $u_{\mathrm{BS}}(\theta)=\exp[\theta(\hat{a}\hat {b}^{\dagger}-\hat a^{\dagger}\hat b)]$ is the beam splitter operator and $\theta$ is the parameter that is related to the reflectivity of beam splitter \cite{Agarwal2012,Kim2002}. Therefore, the output state from the beam splitter can readily be obtained as follows:
   \begin{eqnarray}\label{eqq}
   \ket{\psi}_{\mathrm{out}}&=&\dfrac{1}{\sqrt{2}}\left[ \epsilon_{+}\ket{A}_{\mathrm{out}}\ket{\xi^{'}}_{c}+\epsilon_{+}\ket{B}_{\mathrm{out}}\ket{\xi}_{c}\right.+\left. \epsilon_{-}\ket{C}_{\mathrm{out}}\ket{\xi^{'}}_{c}+\epsilon_{-}\ket{D}_{\mathrm{out}}\ket{\xi}_{c}\right] .
   \end{eqnarray}
   In Eq. (\ref{eqq}) we have defined \footnote[1]{Notice that $u_{\mathrm{BS}}(\theta)\ket{0,0}=\ket{0,0}$ (see Ref. \cite{Agarwal2012}).}:
   \begin{eqnarray}\label{aout}
   \ket{A}_{\mathrm{out}}&=&u_{\mathrm{BS}}(\theta) \ket{\xi}_{a}\ket{-\xi}_{b} =u_{\mathrm{BS}}(\theta)S_{a}(\xi)S_{b}(-\xi)u_{\mathrm{BS}}^{\dagger}(\theta)\ket{0,0}_{a,b},
   \end{eqnarray}
   where for $\theta=\frac{\pi}{4}$ the state (\ref{aout}) results in the well-known two-mode squeezed state as below \cite{Gerry2005}:
   \begin{eqnarray}\label{Twomodesqu}
   \ket{A}_{\mathrm{out}}\rightarrow\ket{\xi}_2&=&S_{\mathrm{ab}}(\xi)\ket{0,0}_{a,b}=\exp(\xi^{\ast}\hat{a}\hat{b}-\xi \hat{a}^{\dagger}\hat{b}^{\dagger})\ket{0,0}_{a,b}\\  \nonumber
   &=&(\cosh{r})^{-1} \Sigma^\infty_{n=0}{(-1)^{n}e^{in\varphi}(\tanh{r})^{n}\ket{n,n}}.
   \end{eqnarray}
   Other constituents of the state ket in (\ref{eqq}) for $\theta=\frac{\pi}{4}$ are defined as:
   \begin{eqnarray}
   \ket{B}_{\mathrm{out}}&=&u_{\mathrm{BS}}(\frac{\pi}{4})\ket{\xi}_{a}\ket{-\xi^{'}}_{b}=u_{\mathrm{BS}}(\frac{\pi}{4})S_{a}(\xi)S_{b}(-\xi)\ket{0,1}_{a,b}\\ \nonumber
   &=&u_{\mathrm{BS}}(\frac{\pi}{4})S_{a}(\xi)S_{b}(-\xi)u_{\mathrm{BS}}^{\dagger}(\frac{\pi}{4})u_{\mathrm{BS}}(\frac{\pi}{4})\hat{b}^{\dagger}u_{\mathrm{BS}}^{\dagger}(\frac{\pi}{4})\ket{0,0}_{a,b}\\ \nonumber
   &=&\dfrac{S_{ab}(\xi)}{\sqrt{2}}\left( \hat{b}^{\dagger}-\hat{a}^{\dagger}\right) \ket{0,0}_{a,b},
   \end{eqnarray}
   \begin{eqnarray}
   \ket{C}_{\mathrm{out}}&=&u_{\mathrm{BS}}(\frac{\pi}{4})\ket{\xi^{'}}_{a}\ket{-\xi}_{b}=u_{\mathrm{BS}}(\frac{\pi}{4})S_{a}(\xi)S_{b}(-\xi)\ket{1,0}_{a,b}\\ \nonumber
   &=&u_{\mathrm{BS}}(\frac{\pi}{4})S_{a}(\xi)S_{b}(-\xi)u_{\mathrm{BS}}^{\dagger}(\frac{\pi}{4})u_{\mathrm{BS}}(\frac{\pi}{4})a^{\dagger}u_{\mathrm{BS}}^{\dagger}(\frac{\pi}{4})\ket{0,0}_{a,b}\\ \nonumber
   &=&\dfrac{S_{ab}(\xi)}{\sqrt{2}}\left( \hat{a}^{\dagger}+\hat{b}^{\dagger}\right) \ket{0,0}_{a,b},
   \end{eqnarray}
   \begin{eqnarray}\label{Components of the total state after the beam splitter}
   \ket{D}_{\mathrm{out}}&=&u_{\mathrm{BS}}(\frac{\pi}{4})\ket{\xi^{'}}_{a}\ket{-\xi^{'}}_{b}=u_{\mathrm{BS}}(\frac{\pi}{4})S_{a}(\xi)S_{b}(-\xi)\ket{1,1}_{a,b}\\ \nonumber
   &=&u_{\mathrm{BS}}(\frac{\pi}{4})S_{a}(\xi)S_{b}(-\xi)u_{\mathrm{BS}}^{\dagger}(\frac{\pi}{4})u_{\mathrm{BS}}(\frac{\pi}{4})a^{\dagger}u_{\mathrm{BS}}^{\dagger}(\frac{\pi}{4})u_{\mathrm{BS}}(\frac{\pi}{4})b^{\dagger}u_{\mathrm{BS}}^{\dagger}(\frac{\pi}{4})\ket{0,0}_{a,b}\\ \nonumber
   &=&\dfrac{S_{ab}(\xi)}{2}\left( \hat{a}^{\dagger}+\hat{b}^{\dagger}\right) \left( \hat{b}^{\dagger}-\hat{a}^{\dagger}\right) \ket{0,0}_{a,b}.
   \end{eqnarray}
   Considering all above results, the output state in (\ref{eqq}) simplifies to
   \begin{eqnarray}\label{total state after the beam splitter}
   \ket{\psi}_{\mathrm{out}}&=&\dfrac{1}{\sqrt{2}}\left[ \epsilon_{+}\ket{\xi}_{2}\ket{\xi^{'}}_{c}+\dfrac{\epsilon_{+}}{\sqrt{2}}S_{ab}(\xi)(\hat{b}^{\dagger}-\hat{a}^{\dagger})\ket{0,0}_{a,b}\ket{\xi}_{c}\right. \\ \nonumber
   &+&\left. \dfrac{\epsilon_{-}}{\sqrt{2}}S_{ab}(\xi)(\hat{b}^{\dagger}+\hat{a}^{\dagger})\ket{0,0}_{a,b}\ket{\xi^{'}}_{c}+\dfrac{\epsilon_{-}}{2}S_{ab}(\xi)(\hat{b}^{\dagger 2}-\hat{a}^{\dagger 2})\ket{0,0}_{a,b}\ket{\xi}_{c}\right] .
   \end{eqnarray}
   Subsequently, Alice performs two photon number measurements on the modes a and b on her side. We denote the probability to find $n$ and $m$ photons respectively in the modes a and b by $P(n,m)$, \textit{i.e.},
   \begin{eqnarray}\label{probability1}
   P(n,m)=\arrowvert_{a}\bra{n}_{b}\bra{m}\psi\rangle_{\mathrm{out}}\arrowvert^2.
   \end{eqnarray}
   In this measurement, if $ n $ and $ m $ differ by one ($ m=n+1 $), the state on Bob's side (the field mode c) collapses into
   \begin{eqnarray}\label{a teleported unknown state}
   \ket{\psi}_{\mathrm{Bob}}=\epsilon_{+}\ket{\xi}_c+\epsilon_{-}\ket{\xi^{'}}_c.
    \end{eqnarray}
    Interestingly, as is clear, provided that the difference between $ n $ and $ m $ being one, our teleportation scheme works perfectly. Under the mentioned conditions, Eq. (\ref{probability1}) reduces to
    \begin{eqnarray}\label{probability2}
    P(n,n+1)=\dfrac{(n+1)\tanh^{2n}r}{4\cosh^4r}.
     \end{eqnarray}
   The success probability to find the number of photons in either mode such that $ m=n+1 $ reads as $P=\Sigma^\infty_{n=0}{P(n,n+1)}$. This summation can be calculated  using (\ref{probability2}), which arrives one at $P=0.25$. This value of success probability is acceptable in the typical schemes for teleportation (see Ref. \cite{Sehati2017,Cardoso2009}).

   \section{The effect of decoherence}
   
   \subsection{Case 1:}
   In this subsection we want to teleport the unknown state (\ref{ateleportedunknownstate}) by using the entangled coherent-squeezed channel,
    \begin{eqnarray}\label{ cha}
    \ket{H_{\xi,\alpha}}^1_{b,c}=\frac{1}{\sqrt{2}}\left( \ket{\alpha}_{b}\ket{\xi^{'}}_{c}+\ket{-\alpha}_{b}\ket{\xi}_{c}\right),
    \end{eqnarray}
    however, under the influence of noise.  In (\ref{ cha}) $ \ket{\pm\alpha} $ are coherent states and $ \ket{\xi} $, $ \ket{\xi^{'}} $ are two squeezed states defined in (\ref{ unknown state}). We assume that the mode b of the maximally entangled channel (\ref{ cha}) suffers from photon loss. This noise is modeled by mixing the state of channel with a vacuum state (we show it by the environmental mode E, $\ket{0}_E$) at a beam splitter with the transmittance of $\eta$ \cite{Van2001,Bergmann2017,Wickert2010}. So, the state (\ref{ cha}) after traveling the mode b from the dissipative channel is converted to the following channel,
    \begin{eqnarray}\label{ chana}
     \ket{H_{\xi,\alpha}}^1_{b,E,c}&=&\frac{1}{\sqrt{2}}\left( \ket{\alpha\sqrt{\eta}}_b\ket{\alpha\sqrt{1-\eta}}_{E}\ket{\xi^{'}}_{c}\right. +\left. \ket{-\alpha\sqrt{\eta}}_b\ket{-\alpha\sqrt{1-\eta}}_{E}\ket{\xi}_{c}\right).
     \end{eqnarray}
    In the continuation, we teleport the unknown state (\ref{ateleportedunknownstate}) using the introduced channel in (\ref{ chana}). So, the whole state reads as,
    \begin{eqnarray}\label{ wholstat}
     \ket{\psi}_a\otimes\ket{H_{\xi,\alpha}}^1_{b,E,c}
     &=&\frac{1}{\sqrt{2}}\left( \epsilon
     _+\ket{\xi,\alpha\sqrt{\eta},\xi^{'}}_{a,b,c}\ket{\alpha\sqrt{1-\eta}}_E\right.+\left.\epsilon
     _+\ket{\xi,-\alpha\sqrt{\eta},\xi}_{a,b,c}\ket{-\alpha\sqrt{1-\eta}}_E\right.\\\nonumber
     &+&\left. \epsilon
     _-\ket{\xi^{'},\alpha\sqrt{\eta},\xi^{'}}_{a,b,c}\ket{\alpha\sqrt{1-\eta}}_E+\epsilon
     _-\ket{\xi^{'},-\alpha\sqrt{\eta},\xi}_{a,b,c}\ket{-\alpha\sqrt{1-\eta}}_E\right).
     \end{eqnarray}
     By operating a proper measurement of the quasi-Bell state, \textit{i.e.}, $ |qB\rangle _{ab}\langle{qB}|$, where,
     \begin{eqnarray}\label{ quasiBell}
       \ket{qB}_{a,b}=\frac{1}{\sqrt{2}}\left( \ket{\xi,-\alpha\sqrt{\eta}}+\ket{\xi^{'},\alpha\sqrt{\eta}}\right),
       \end{eqnarray}  
  on the state (\ref{ wholstat}) and then, by choosing $ \alpha $ and $ \eta $ in such a way that $ 2\eta|\alpha|^{2} $ is enough large such that the terms consist of $ e^{-2\eta|\alpha|^{2}} $ can be ignored, the state of Bob may be approximated to the state (\ref{ateleportedunknownstate}) as follows,
  \begin{eqnarray}\label{case1}
  \ket{\psi}^1_{\mathrm{Bob}}=\epsilon_{+}\ket{\xi}_c\ket{-\alpha\sqrt{1-\eta}}_E+\epsilon_{-}\ket{\xi^{'}}_c\ket{\alpha\sqrt{1-\eta}}_E.
   \end{eqnarray}
  In this way, our teleportation protocol is near to perfect teleportation (see Refs. \cite{Van2001,Cardoso2009}) with the fidelity,
   \begin{eqnarray}\label{ fidelitya}
        F=\left| _{a}\langle\psi|\psi\rangle^1_{\mathrm{Bob}}\right|^2=|\epsilon_{+}|^{4}+|\epsilon_{-}|^{4}+2|\epsilon_{+}|^{2}|\epsilon_{-}|^{2}e^{-2(1-\eta)|\alpha|^{2}}.
        \end{eqnarray}
        In the following, the fidelity, F, has been plotted in figure \ref{fig:Fig2} for different chosen values of $ \alpha $ and $ \eta $ versus $ |\epsilon_{-}|^{2} $ (the unknown coefficient of $ \ket{\xi^{'}} $ in state (\ref{ateleportedunknownstate})). In this figure, solid red line corresponds to $ \eta=1 $ (absence of decoherence) where as we expect fidelity is equal to one; \textit{i.e.}, we have perfect teleportation. In the presence of decoherence $(\eta\neq 1)$ by decreasing $ \eta $ in this figure, the minimum of the fidelity has been decreased. Also, from the initial and final points of the curves in figure \ref{fig:Fig2}, we notice that when we have only the net state of $ \ket{\xi} $ or $ \ket{\xi^{'}} $ in state (\ref{ateleportedunknownstate}), fidelity is equal to one even in the presence of decoherence and we have again perfect teleportation.
         \begin{figure}[H]
           \centering
         \includegraphics[width=0.75\textwidth] {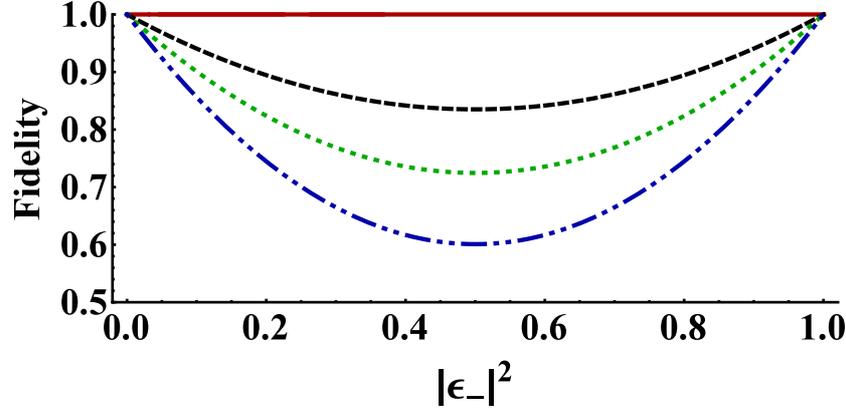}
           \caption{\label{fig:Fig2} {Fidelity in Eq. (\ref{ fidelitya}) versus the amplitude of probability, for $ \eta=1 $ (solid red line),  $ \eta=0.95 $ (dashed black line), $ \eta=0.9 $ (dotted green line) and $ \eta=0.8 $ (dot-dot-dashed blue line) with $ \alpha=2 $.}}
          \end{figure}
   
  \subsection{Case 2:}
   In this subsection we want to teleport the following unknown state,
  \begin{eqnarray}\label{state2}
  \ket{\psi^{'}}_a=A(\xi)\left( a_+\ket{\xi}_{a}+a_-\ket{-\xi}_{a}\right),
  \end{eqnarray}
  where $ \ket{\pm\xi} $ are two squeezed vacuum, $a_\pm$ are unknown expansion coefficients and
  \begin{eqnarray}
  A(\xi)=\left(\left|a_+ \right|^2+ \left|a_- \right|^2+ (a^{*}_+a_-+a_+a^{*}_-) \sqrt{\frac{1-\tanh^2r}{1+\tanh^2r}}\right) ^{-\frac{1}{2}},
  \end{eqnarray}
  using the entangled coherent-squeezed channel,
  \begin{eqnarray}\label{chan}
  \ket{H_{\xi,\alpha}}^2_{b,c}=M(\xi,\alpha)\left( \ket{\alpha}_{b}\ket{-\xi}_{c}+\ket{-\alpha}_{b}\ket{\xi}_{c}\right),
  \end{eqnarray}
  in the presence of decoherence, where $ \ket{\pm\alpha} $ are coherent states and
  \begin{eqnarray}
  M(\xi,\alpha)=\frac{1}{\sqrt{2}}\left( 1+e^{-2\left| \alpha\right|^2 } \sqrt{\frac{1-\tanh^2r}{1+\tanh^2r}}\right) ^{-\frac{1}{2}}.
  \end{eqnarray}
  The entangled channel (\ref{chan}) can be produced in the resonant
  interaction of a $\Lambda$-type atom with a two-mode quantized field in the presence of two strong classical fields \cite{Karimi2016}. Also, we consider the photon loss for mode b of the channel (\ref{chan}). So, the state (\ref{chan}) after traveling the mode b from dissipative channel is converted to:
  \begin{eqnarray}\label{dissstate}
  \ket{H_{\xi,\alpha}}^2_{b,E,c}&=&M(\xi,\alpha)\left( \ket{\alpha\sqrt{\eta}}_b\ket{\alpha\sqrt{1-\eta}}_{E}\ket{-\xi}_{c}\right. +\left. \ket{-\alpha\sqrt{\eta}}_b\ket{-\alpha\sqrt{1-\eta}}_{E}\ket{\xi}_{c}\right).
  \end{eqnarray}
   Now, we want to teleport the unknown state (\ref{state2}) using the dissipative channel (\ref{dissstate}). The whole state then reads as
  \begin{eqnarray}\label{eq27}
  \ket{\psi^{'}}_a\otimes\ket{H_{\xi,\alpha}}^2_{b,E,c}
  &=& M(\xi,\alpha)A(\xi)\left( a
  _+\ket{\xi,\alpha\sqrt{\eta},-\xi}_{a,b,c}\ket{\alpha\sqrt{1-\eta}}_E\right.\\\nonumber
  &+& a
  _+\ket{\xi,-\alpha\sqrt{\eta},\xi}_{a,b,c}\ket{-\alpha\sqrt{1-\eta}}_E+ a
  _-\ket{-\xi,\alpha\sqrt{\eta},-\xi}_{a,b,c}\ket{\alpha\sqrt{1-\eta}}_E\\\nonumber
  &+&\left. a
  _-\ket{-\xi,-\alpha\sqrt{\eta},\xi}_{a,b,c}\ket{-\alpha\sqrt{1-\eta}}_E\right) .
  \end{eqnarray}
   By operating a proper measurement using the quasi-Bell state,
     \begin{eqnarray}\label{ quasiBellb}
       \ket{qB}_{ab}=C(\alpha,\xi)\left( \ket{\xi,-\alpha\sqrt{\eta}}+\ket{-\xi,\alpha\sqrt{\eta}}\right),
       \end{eqnarray}  
    where $ C(\alpha,\xi) $ is the normalization coefficient, via the projection operator $ |qB\rangle_{ab}\langle{qB}| $ on the state (\ref{eq27}) and then choosing $ \alpha $, $ \eta $ and $ r $ in such a way that the terms consist of $ e^{-2\eta|\alpha|^{2}} $ and $ \dfrac{1}{\sqrt{2\cosh^{2}r-1}} $ in the outcome state can be ignored, the Bob's state then reads as
    \begin{eqnarray}\label{ bobstat}
         \ket{\psi}^2_{\mathrm{Bob}}=A(\xi,\alpha)\left(a_+\ket{\xi}_{c}\ket{-\alpha\sqrt{1-\eta}}_{E}+a_-\ket{-\xi}_{c}\ket{\alpha\sqrt{1-\eta}}_{E}\right),
         \end{eqnarray}
     where
     \begin{eqnarray}\label{ coficiant}
            A(\xi,\alpha)=\left(|a_+|^{2}+|a_-|^{2}+(a^{*}_{+}a_{-}+a_{+}a_{-}^{*})e^{-2(1-\eta)|\alpha|^{2}}\sqrt{\dfrac{1-\tanh^{2}r}{1+\tanh^{2}r}}\right)^{-\dfrac{1}{2}}.
            \end{eqnarray}
     As is seen, the state of Bob, Eq. (\ref{ bobstat}), is near to state (\ref{state2}) and so we achieved to our teleportation purpose near to perfect teleportation with the fidelity that may be obtained via, 
   \begin{eqnarray}\label{fideltiyb}
   F&=&\left| _{a}\langle\psi'|\psi\rangle^2_{\mathrm{Bob}}\right|^2\\\nonumber
   &=&|A(\xi,\alpha)|^{2}|A(\xi)|^{2}\left[ | a
    _+|^{4}+| a
      _-|^{4}+2(| a_+|^{2}+| a_-|^{2})Re(a_{+}^{*}a_{-})\langle \xi \arrowvert -\xi\rangle e^{-2(1-\eta)|\alpha|^{2}}\right.\\\nonumber
   &+& 2|a_+|^{2}| a_-|^{2}(|\langle \xi \arrowvert -\xi\rangle|^{2}+e^{-2(1-\eta)|\alpha|^{2}})+2(|a_+|^{2}+|a_-|^{2})Re(a_{+}^{*}a_{-})\langle \xi \arrowvert -\xi\rangle\\\nonumber
      &+&\left. |\langle \xi \arrowvert -\xi\rangle|^{2}e^{-2(1-\eta)|\alpha|^{2}}(a_{+}^{*2}a_{-}^{2}+a_{-}^{*2}a_{+}^{2})\right]  .
   \end{eqnarray}
        To plot the fidelity using Eq. (\ref{fideltiyb}), we consider $ a_{+}=\cos\theta $ and $ a_{-}=\sin\theta $. Fidelity, $F$, is plotted for different  $ \eta $ versus $ \theta $ when $ \alpha=2 $ and $ r=7 $ are fixed (see figure \ref{fig:Fig3}). Solid red line is plotted for $ \eta=1 $ (absence of decoherence) and as we expect fidelity is equal to $ 1 $, \textit{i.e.}, we have perfect teleportation. However, in this figure, in the presence of decoherence $(\eta\neq 1)$, by decreasing $ \eta $, the minimum of  fidelity has been decreased.
  \begin{figure}[H]
           \centering
         \includegraphics[width=0.75\textwidth] {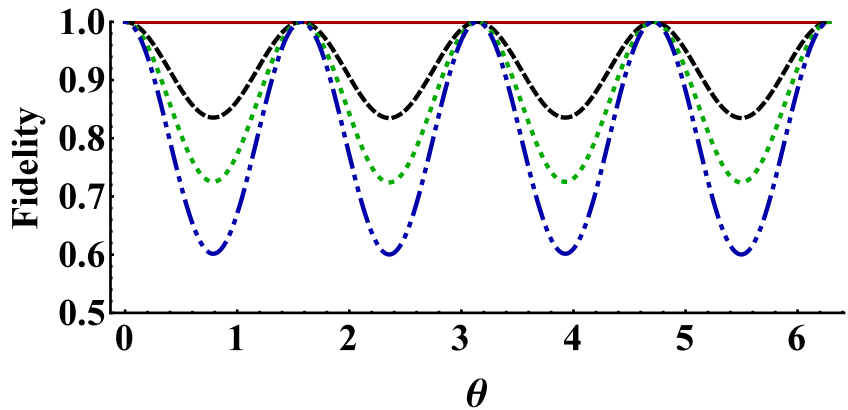}
           \caption{\label{fig:Fig3} {Fidelity in Eq. (\ref{fideltiyb}), for $ \eta=1 $ (solid red line), $ \eta=0.95 $ (dashed black line), $ \eta=0.9 $ (dotted green line) and $ \eta=0.8 $ (dot-dot-dashed blue line) with $ \alpha=2 $ and $ r=7 $.  }}
          \end{figure}      
\section{Summary and conclusions} \label{sec.Conclusion}
In this paper we investigated teleportation of two different superpositions of squeezed states in the absence and presence of dissipation. According to our analytical results, in the absence of photonic loss, the unknown state which includes a superposition of squeezed vacuum and squeezed one-photon states was teleported with fidelity (success probability) as $1$ ($0.25$). Moreover, under the influence of dissipation, the same unknown state was teleported near perfectly. In addition, we considered the teleportation of a superposition of two squeezed vacuum states separated in phase by $180^\circ$. This superposition under proper conditions leads to perfect teleportation (recently, approximate  and near to perfect teleportation are reported \cite{Van2001,Cardoso2009}). Also, we found that in the presence of decoherence $(\eta\neq 1)$, the minimum of fidelity is decreased by decreasing $\eta$. Summing up, we could successfully teleport the considered superpositions of squeezed states by properly chosen squeezed entangled state and coherent-squeezed entangled  channels with utilizing a beam splitter and applying proper measurements while some suitable conditions are fixed.


\end{document}